\documentclass[prc,12pt,english]{revtex4}
\usepackage{babel}
\usepackage{amsmath}
\usepackage{graphics}
\usepackage{pdfpages}

\newcommand\nn{\nonumber}
\newcommand\ba{\begin{eqnarray}}
\newcommand\ea{\end{eqnarray}}
\newcommand\be{\begin{equation}}
\newcommand\ee{\end{equation}}
\newcommand\alb{\begin{align}}
\newcommand\ale{\end{align}}

\usepackage{float}
\usepackage{amsmath} 
\usepackage[title]{appendix}

\begin{document}

\title{Modeling the effects of natural disasters, wars, and migrations on sustainability or collapse of pre-industrial societies: Random perturbations of the Human and Nature Dynamics (HANDY) model}

\author{Lo\"ic Patry}
\email{loic.patry@universite.paris-saclay.fr}
\affiliation{\it Université Paris-Saclay, 91191
Gif-sur-Yvette Cedex, France}

\author{Pierre Morel}
\email{pierre.morel@lpp.polytechnique.fr}
\affiliation{ Laboratoire de Physique des Plasmas, CNRS, Universit\'e Paris-Saclay, \'Ecole Polytechnique, 91128, Palaiseau Cedex}

\author{Egle~Tomasi-Gustafsson} \email{egle.tomasi@cea.fr}
\affiliation{\it DPhN, IRFU, CEA, Universit\'e Paris-Saclay, F-91191
Gif-sur-Yvette Cedex, France}

\author{Eugenia~Kalnay} \email{ekalnay@umd.edu}
\affiliation{\it Department of Atmospheric and Oceanic Science, University of Maryland, College Park, Maryland, 20742, USA}

\author{Jorge Rivas}
\email{jorgerodrigorivas@gmail.com}
\affiliation{\it Independent Researcher}

\author{Safa~Mote} \email{mote@pdx.edu}
\affiliation{\it Fariborz Maseeh Department of Mathematics and Statistics, Portland State University, Portland, Oregon, 97201, USA; \\
Department of Atmospheric and Oceanic Science, University of Maryland, College Park, Maryland, 20742, USA}

\begin{abstract}

We study the effect of random perturbations in the Human and Nature Dynamics (HANDY) model. HANDY models the interactions between human population, depletion, and consumption of natural resources. HANDY explains how endogenous human--nature interactions could lead to sustainability or collapse in past societies. We introduce a Gaussian random noise perturbation on the population change to represent generic external perturbations. The robustness of the results is investigated with statistical analysis based on probability distributions of specific events. Our study shows that the results of the unperturbed HANDY model are robust under small perturbations of $\lesssim$ 10\% of the Human population. Our results confirm that endogenous dynamics drive the societal cycles. However, exogenous perturbations, such as floods, droughts, earthquakes, volcanic eruptions, infectious disease, epidemics, and wars, can accelerate or delay a collapse cycle.
\end{abstract}
\date{\today}
\maketitle

\section{Introduction}

Recent decades have seen an increased awareness about the fragility of nature due to the actions of humans. Objective data and projections from the scientific community on climate change \cite{GIEC} as well as more frequent climatic catastrophes that affect directly or indirectly a larger fraction of the population increase the conviction that it is urgent to take action. The rapid increase of human population2, which has surpassed 8 billion, combined with a similarly rapid increase in consumption \textit{per capita}, have resulted in a remarkable growth of the total human impact on Earth. This growth has led to a plethora of effects on the Earth system, such as loss of biodiversity,  soil erosion, and other  factors that may drive Earth subsystems to collapse \cite{lenton_PNAS_2008, motesharrei_modeling_2016, carrying_capacity_mote_2020}.

Historians, archeologists, and increasingly natural scientists and other scholars, have offered numerous explanations for the many individual cases of societal collapses in history. Particular explanations for each of the cases of collapse in history include one or more of the following causes: natural causes such as droughts, soil erosion or exhaustion, salinization of soils, deforestation, changes in climatic conditions, depletion of particular mineral resources, volcanoes, earthquakes, floods, and changes in the courses of rivers; as well as human causes such as foreign invasions, tribal and societal migrations, popular uprisings, civil wars, changes in trade patterns, cultural decline or social decadence, and technological changes such as the introduction of ironworking, the introduction of wheeled wagons, or advancements in the methods and arms of warfare, such as the introduction of horse cavalry, armored infantry, long swords, or firearms. In the Human and Nature Dynamics (HANDY) paper \cite{MOTESHARREI201490} \cite{MOTESHARREI201490}, it was argued that the fundamental scientific uncertainties underlying these kinds of very specific theories are that each cause is particular to that specific instance of societal collapse, and even in the cases where they seem to strongly contribute to that case of collapse, that specific society had almost always already previously experienced that particular cause without collapsing. The paper also mentioned the  well-known example of Minoan civilization, which had experienced earthquakes that had destroyed palaces and cities numerous times, yet they would just rebuild even more grandly. In fact, droughts, floods, volcanoes, earthquakes, soil erosion, deforestation and climate changes have been common in history without inducing societal collapses. Similarly, foreign invasions, human migrations, popular uprisings, and civil wars have occurred repeatedly throughout human history without inducing societal collapses, including in the cases where they did eventually contribute to collapse. The most prominent cases of the Babylonian, Assyrian, Mauryan, Han, and Roman empires effectively warred against and held off, often for centuries, the nearby rivals or “barbarian” peoples, but who then ultimately defeated the empire contributing to a societal collapse.

With regards to both natural and human causes of collapse, actual specific collapses may have been due to a number of specific causes, but each particular explanation produces the scientific uncertainty of why this specific occurrence of this cause induced a collapse when in other previous occurrences it did not. Thus, the original HANDY paper argued that while individual causes such as those identified in theories of specific collapses can play a role in those specific historical cases, the widespread existence of collapse across the societies of the world, and across human history, argues for a more general explanation than just these particular causes of each individual case:  the HANDY model suggests the long term evolution and interaction of the over-depletion of nature and the over-exploitation of labor as the underlying critical causes.  However, this does not mean that individual causes such as those outlined above do not play any role in  societal evolution and potential collapse.

HANDY is a minimal dynamical system that describes the interaction between humans and nature \cite{MOTESHARREI201490}. Although HANDY was not the first study of this kind (see \cite{BranderTaylor1998} and references therein), HANDY modeled accumulation of wealth and inequality for the first time with a relatively small number of equations and parameters. The model is based on a system of four interconnected equations of Lotka--Volterra type \cite{Lotka1910,volterra}, and describe the evolution of pre-industrial human population, divided into two groups, the Commoners and Elites, coupled with the accumulated Wealth and Nature. The Lotka--Volterra equations, which model the 'predator--prey' systems, were originally developed to describe fish populations in the Adriatic Sea \cite{volterra}. Since then, these equations have been applied to many different domains, for example,  nonlinear optics \cite{photonics9010016}, genetics \cite{PhysRevE.96.022416}, population dynamics \cite{Monte2009PredictingTE}, and plasma physics \cite{Morel_2013}.

This work aims to investigate the stability of the solutions of HANDY under small and large perturbations. Different integrators and potential sources of instabilities are studied. As the equations must be integrated numerically, instabilities that bring specific configurations to collapse may  relate to the integration method and its parameters. Moreover, a random noise is added to one equation. In particular, large sudden fluctuations in population may occur in the real world due to wars, famines, pandemics, or other natural disasters.
The solutions are collected and analyzed statistically based on a large number of numerical experiments. The results are discussed in terms of probability of collapse. Special attention is paid to the possible deviations of the populations from a Gaussian distribution in terms of skewness and kurtosis.

\subsection{A brief description of the HANDY model}

The extension of the Lotka-Volterra equations, presented in HANDY, is interesting for several reasons. HANDY deals with a complex problem: the coupled evolution of human population, natural resources, wealth, and their interactions are minimally described with variables that simultaneously take into account several effects.
HANDY applies to societies after the Agricultural Revolution and before the Industrial Revolution (i.e., the exploitation of fossil fuels). HANDY results show that economic stratification or over-depletion of Nature can jointly, or each independently, lead to population collapse. An extension to industrial societies is currently being developed.

Four coupled equations describe the evolution of (a) "the Commoners", $x_C$:  the population that labors to produce Wealth, (b) "the Elites", $x_E$: the population that controls Wealth and can therefore consume a larger share of Wealth, (c) Nature, $y$,  representing regenerating natural resources, and d) the accumulated Wealth, $w$:
\be
\begin{array}{lccl}
 \textnormal{a) \ Commoners: \,\,\,\,\,}  & \displaystyle\frac{dx_C}{dt} & = & \beta_C x_C-\alpha_C(x_C,w)x_C , \nn \\
 \textnormal{b) \ Elites:} & \displaystyle \frac{dx_E}{dt} & = & \beta_E x_E-\alpha_E(x_E,w)x_E ,\nn \\
 \textnormal{c) \ Nature: } & \displaystyle\frac{dy}{dt} & = & \gamma y(\lambda-y)-\delta x_Cy, \nn\\
 \textnormal{d) \ Wealth: } & \displaystyle\frac{dw}{dt} & = & \delta x_Cy-C_C(x_C,w)-C_E(x_E,w). \nn
\end{array}
\label{eq:Handy}
\ee
The separation of population into Commoners and Elites models inequality and economic stratification. For these two populations, there is a birth rate ($\beta$) with a fixed value and a death rate ($\alpha$) that increases if Wealth falls below a certain threshold.
 Famine starts when consumption, $i.e.$, the rate at which Commoners or Elites are using Wealth, drops below a certain threshold.
The death rates for Commoners, $\alpha_C$  and Elites  $\alpha_E$ are given by:
\begin{eqnarray}
 \alpha_C & = &
 \alpha_m + \textnormal{max} \left[ 0,  \left (1 - \displaystyle \frac{C_C}{s x_C} \right ) \right ]
 \left ( \alpha_M - \alpha_m \right ),  \nn\\
 \alpha_E & = &
 \alpha_m + \textnormal{max} \left [ 0,  \left ( 1 - \displaystyle \frac{C_E}{s x_E} \right ) \right ]
 \left ( \alpha_M - \alpha_m \right ).
\label{eq:alpha}
\end{eqnarray}
$\alpha_C$  and $\alpha_E$ vary between  $\alpha_m$ and $\alpha_M$, which are parameters indicating respectively the normal death rate and the famine death rate.
The Nature equation contains a regeneration term, which is a logistic function saturating at the level of $\lambda$, the Nature’s Capacity. Nature is depleted at a rate  $\delta x_Cy$, which is the same as the production of Wealth. The model assumes no losses from depletion of natural resources to production of Wealth.

Finally, $C_C$ and $C_E$ are the human consumption terms for Commoners and Elites, respectively, where $s$ stands for the subsistence salary \textit{per} \textit{capita}.  The different levels of consumption between Commoners $C_C$ and Elites $C_E$ is included as a multiplicative factor $\kappa $, which would be equal to 1 in an equitable society and larger than 1 otherwise.
\begin{eqnarray}
 \label{eq:eqCC}
 C_C &=& \textnormal{min}\left (1,\displaystyle\frac{w}{w_{th}}\right ) sx_C, \nn \\
 C_E &= &\textnormal{min}\left (1,\displaystyle\frac{w}{w_{th}}\right )\kappa sx_E,
\label{eq:eqCE}
\end{eqnarray}
where $w_{th}=\rho x_C +\kappa \rho x_E$ is a threshold level for Wealth and $\rho$ is the threshold Wealth per capita.

For simplicity, we limit our analysis to an egalitarian society without Elites. Therefore, the distinction between Elites and Commoners disappears and they are indistinctly referred to as Humans in the rest of this paper.
\section{Study of instabilities}
\subsection{The integration method}

The numerical integration method may influence  the results. The principle of numerical integration is to discretize the variable $x$ as a function of time, then to calculate the value $x_{n+1}$  at $t+1$ from $x_{n}$ at time $t$. In some calculations, due to discretization, $y$ can become negative, which then leads to a divergence towards $- \infty$ because of the structure of the logistic equation. It is therefore necessary to constrain $y$ to zero as soon as it becomes negative. An additional constraint, $x_C = 0$,  is set when the human population becomes smaller than two units $x_C \le 2$.



\subsection{Replacement of $C_C$ and $C_E$ by smooth functions}
\begin{figure}[h]
\centering
\includegraphics[width=0.90\linewidth]{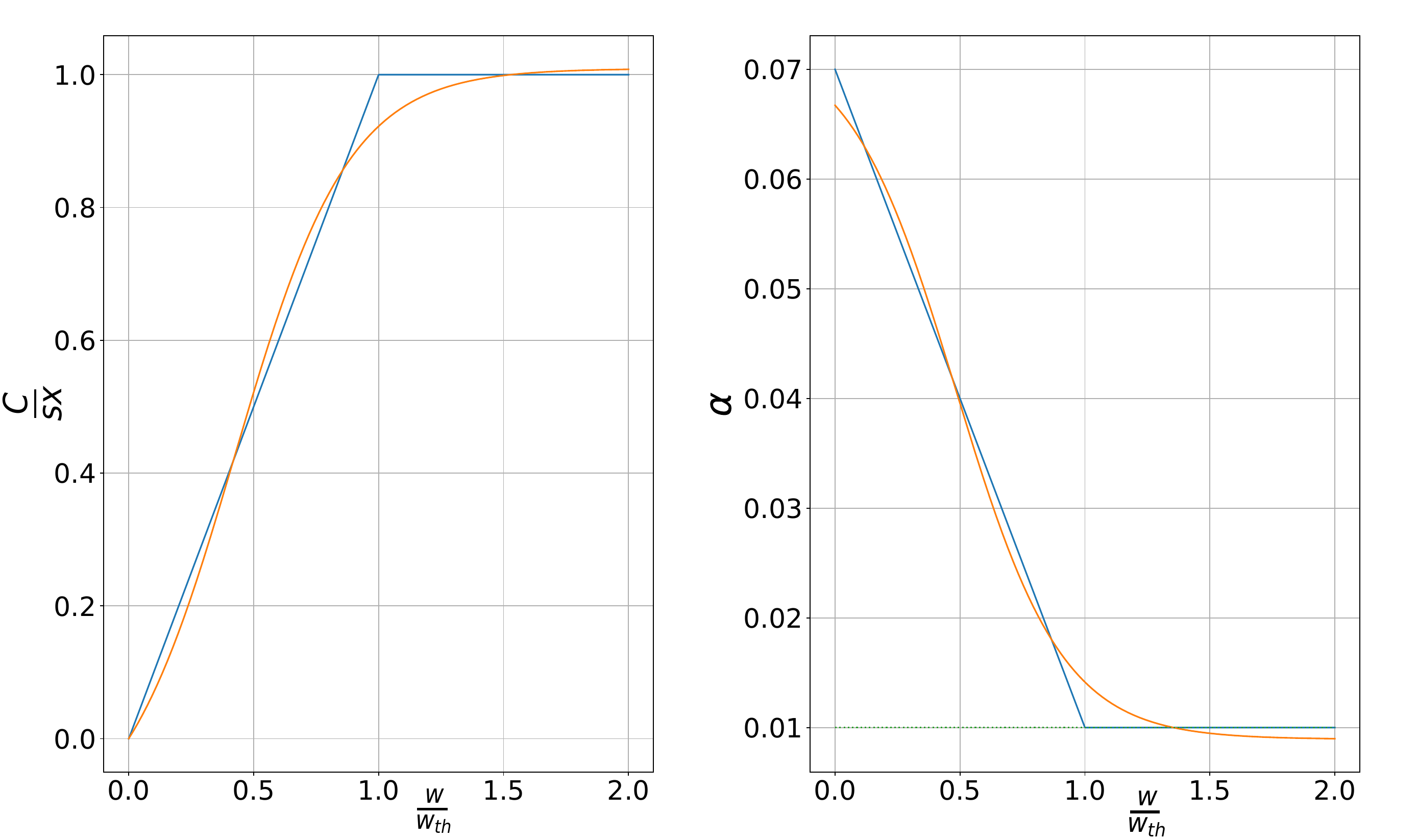} \\
\caption{Superposition of the two parametrizations of $C$ (left) and $\alpha$(right)  (see text). }
\label{Fig:calpha}
\end{figure}

In Ref. \cite{MOTESHARREI201490}, the expressions for $\alpha$ and $C$ are not differentiable because of the min() and max() functions. To check if such a parametrization is a source of instabilities, the expressions(\ref{eq:alpha}) have been replaced by differentiable functions of the form
\be
 y=\displaystyle\frac{A}{1+exp\left(\displaystyle\frac{x-B}{C}\right)}+ D,
\label{eq:eqCF}
\ee
with parameters set by a minimization procedure,  as provided in table \ref{tab:my_label}.
\begin{table}[h]
    \centering
    \begin{tabular}{|c|c|c|c|c|}
    \hline
    &A& B& C& D \\
    \hline
    $\displaystyle\frac{C}{sx}$ & 1.16686& 0.423961& -0.228572& -0.157721 \\
    $\alpha$ & 0.0638524& 0.482244 &0.213897 &0.00893366\\
    \hline
    \end{tabular}
    \caption{Values of the parameterizations of $C$ and $\alpha$.}
    \label{tab:my_label}
\end{table}
The two superposed curves are visualized in Fig. \ref{Fig:calpha}.\
After introducing these differentiable functions into the model equations, the nature of the results (oscillations, equilibrium, collapse) is not changed, but in some cases the amplitudes of the curves are slightly affected.

\subsection{Test of a Gaussian background noise}
In an experiment,  the key signal is often perturbed by other phenomena of lower intensity. To model this effect and to test the stability of HANDY against such effects which could be seen as natural disasters, an additive noise of Gaussian (or normal) distribution is added to the population term. 

The differential equation for  the Human Population (Commoners) are modified as:
\begin{equation}
    \frac{dx_C}{dt} = \beta x_C - \alpha  x_C
    \label{Eq:eqx}
\end{equation}
into:
\begin{equation}
    \frac{dx_C}{dt} = \beta x_C - \alpha  x_C + \epsilon R x_C,
      \label{Eq:eqx1}
\end{equation}
{\it i.e.,} by adding a term $\epsilon  R x_C$, where $R$ is a random number chosen from a normal distribution centered in 0 with a standard deviation $\sigma = 1$, and $\epsilon $ is chosen in the interval $0\le \epsilon \le 1 $. 
A new random number $R$ is chosen after each time interval $\Delta t = 1$ year. Even with an integration time step smaller than one year, we keep $R$ the same over that year so that we have the feature of good years and bad years. Otherwise, rapidly fluctuating perturbations can cancel each other.

\begin{figure}[H]
\centering
\includegraphics[scale=1]{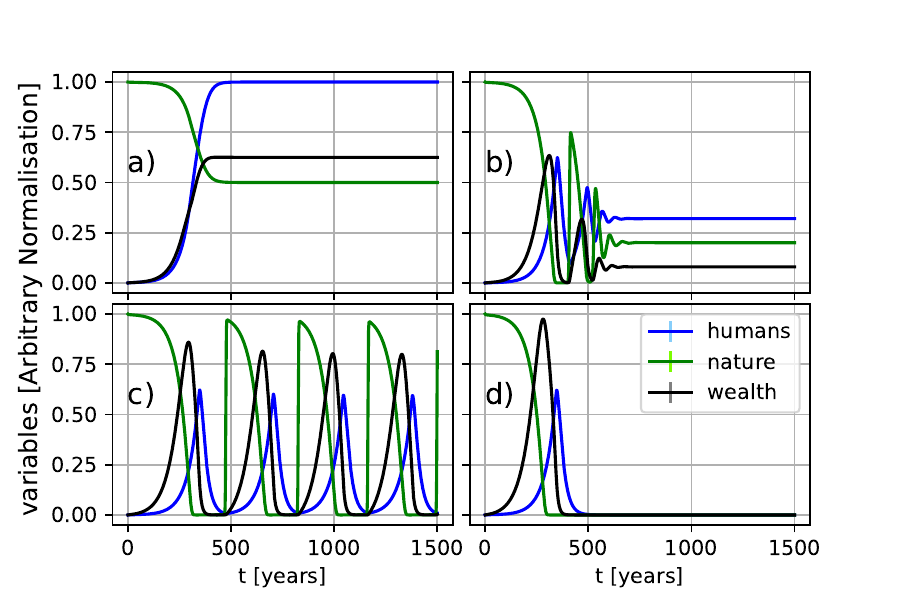} \\
\caption{Time evolution of Nature (green line), Wealth (black line) and Humans (blue line) in an egalitarian society, for a) $\delta=\delta_{opt}$: b) $\delta=2.5\delta_{opt}$:  c) $\delta=4\delta_{opt}$; d)  for $\delta=5.5 \delta_{opt}$. The other parameters are as in Ref. \cite{MOTESHARREI201490}.}
\label{fig:eps0}
\end{figure}

As a reference case, the results of Ref. \cite{MOTESHARREI201490} are reproduced in Fig. \ref{fig:eps0} in the case of an egalitarian society with  $\epsilon =0 $.

Fig. \ref{fig:eps0} reproduces the four basic scenarios as in  Ref. \cite{MOTESHARREI201490}, corresponding to four values of the depletion factor $\delta = \left \{ 1 ; 2.5 ; 4 ; 5.5 \right \} \delta_{\textnormal{\tiny opt}}$. In Fig. \ref{fig:eps0}a, at the optimal value $\delta = \delta_{\textnormal{\tiny opt}}$, the state variables go to equilibrium, while in Fig. \ref{fig:eps0}b they partially overshoot and then oscillate towards equilibrium. The structure is different in Fig. \ref{fig:eps0}c: $\delta$ is too high and Nature falls close to zero. Thus the Human population also collapses and this cycle starts over.
If $\delta$ is increased even more, Nature goes to zero, and once at zero it cannot grow back again due to the nature of the logistic equation. This scenario shows a full collapse of the system and is called \emph{Type-N} collapse, where \emph{N} denotes Nature. This scenario has a plausible real life explanation: if Nature fully collapses, then it would take  a very long time for Nature to thrive again.
\section{Amplitude of the random noise}
\label{sec:ampl}
\begin{figure}[H]
 \centering
 \includegraphics[scale=1]{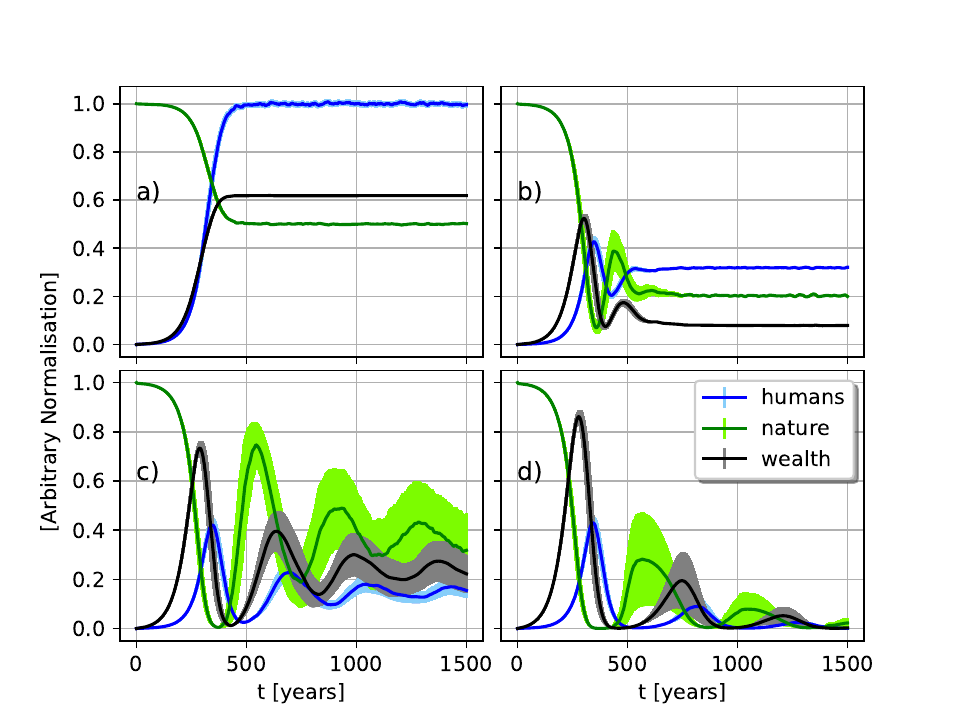}
 \caption{Time evolution of the perturbed system over 1500 years  as in Fig. \ref{fig:eps0},  adding a random noise of amplitude $\epsilon$ = 0.03 that follows a Gaussian distribution. The standard deviation of the set of 1000 trials is visualized as an ensemble spread around a solid line representing the ensemble mean.}
 \label{fig:eps0.03}
\end{figure}

After substituting  Eq. (\ref{Eq:eqx}) by Eq. (\ref{Eq:eqx1}), the equations are integrated numerically for different values of $\epsilon$ in the interval  $0<\epsilon<1$. For each value of $\delta$ and $\epsilon$, 1000 trials are  computed,
The results of  the simulation, repeated 1000 times with the amplitude of the random noise $\epsilon$ fixed at 0.03, are illustrated in Fig. \ref{fig:eps0.03}. The values at each time-step are averaged and their standard-deviation is computed. The results are plotted as a function of time similar to Fig. \ref{fig:eps0}. The solutions are shown as a spread representing the standard deviation of the set of trials around the  average distribution.

Apart from a wider spread of the distribution, Figs. \ref{fig:eps0.03}a and \ref{fig:eps0.03}b do not show a noticeable difference with respect to the reference case  in Figs. \ref{fig:eps0}a and \ref{fig:eps0}b. The state variables  evolve and oscillate towards a steady state.
In Fig. \ref{fig:eps0.03}c, the amplitude decreases in time whereas the corresponding reference case shows cycles of similar amplitude. The widening of the distribution becomes more visible in this figure. Lastly in Fig. \ref{fig:eps0.03}d, two additional oscillations appear, compared to the original version where the collapse is reached directly. In general, adding random noise marginally modifies the dynamics of $\delta$ close to $\delta_{\textnormal{\tiny opt}}$, while it does affect more strongly the higher values: either damping the oscillatory case or preventing some cases from collapse.

The dispersion between the realizations, as encoded by the standard deviation, becomes larger for the two highest values of the depletion factor, $\delta = \{ 4.0, 5.5 \}\, \delta_{\textnormal{\tiny opt}}$ (Figs. \ref{fig:eps0.03}c, \ref{fig:eps0.03}d).

\begin{figure}[H]
\centering
\includegraphics[scale=1]{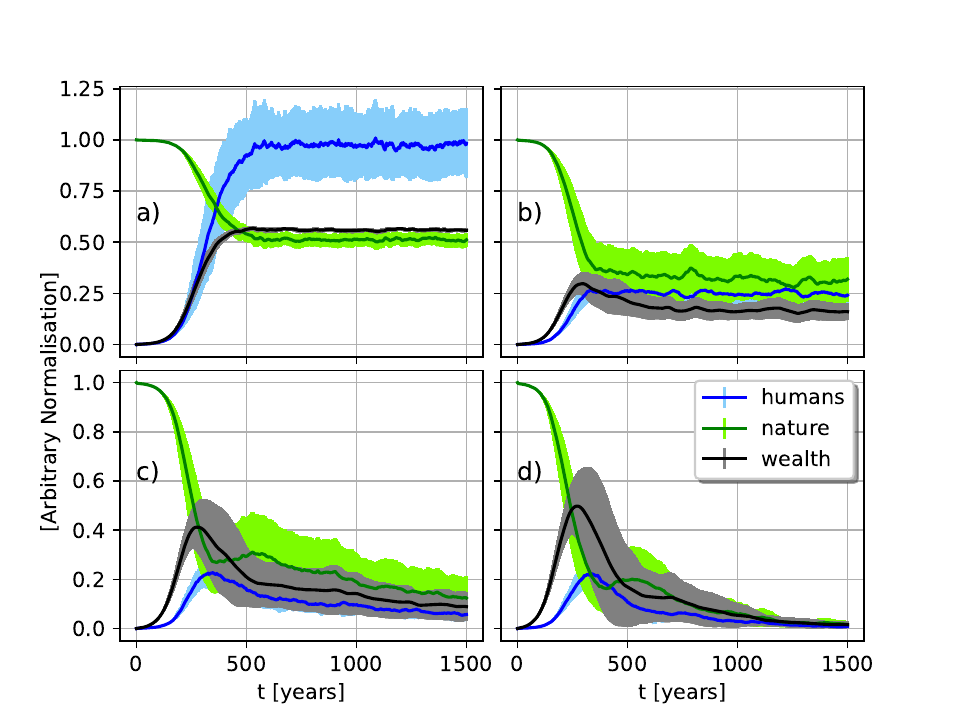} \\
\caption{Same as Fig. \ref{fig:eps0.03} but for $\epsilon$ = 0.1.}
\label{fig:eps0.1}
\end{figure}

\noindent Fig. \ref{fig:eps0.1} shows the results of a similar procedure, but with the amplitude of the random noise $\epsilon$ fixed at 0.1.
\\
In Fig. \ref{fig:eps0.1}a, the global trend does not change compared to the reference version but the standard deviation is larger than for the previous cases. In Fig.  \ref{fig:eps0.1}b, the oscillations preceding the steady state are dampened to the extent that they disappear. At $\epsilon \ge 0.1$, all oscillations disappear; the system either evolves to steady state \ref{fig:eps0.1}a-b or to collapse \ref{fig:eps0.1}c-d. The model  seems to have lost its oscillatory characteristics. This is expected since for most time steps the magnitude of the exogenous perturbation becomes larger than the endogenous change in state variables.

\section{Departure from gaussianity}
The skewness measures the degree of asymmetry of a given set of realizations, and is defined as the third moment of a distribution:
\begin{equation}
 \mathcal{S}_x = \frac{\overline{\left ( x_i - \overline{x_i} \right )^3}}{\left [ \overline{\left ( x_i - \overline{x_i} \right )^2} \right ]^{3/2}} \, ,
 \label{eq:skewness}
\end{equation} where $x_i$ represents the set of realizations simulated for the variable $x$, and $\overline{x_i}$ stands for the average over the realizations.

The departure from a Gaussian distribution can be quantified by the fourth order moment of a distribution, the kurtosis, which is defined as:
\begin{equation}
 \mathcal{K}_x = \frac{\overline{\left ( x_i - \overline{x_i} \right )^4}}{\left [ \overline{\left ( x_i - \overline{x_i} \right )^2} \right ]^{2}} \, .
 \label{eq:kurtosis}
\end{equation}

\begin{figure}[H]
\centering
\includegraphics[scale=1]{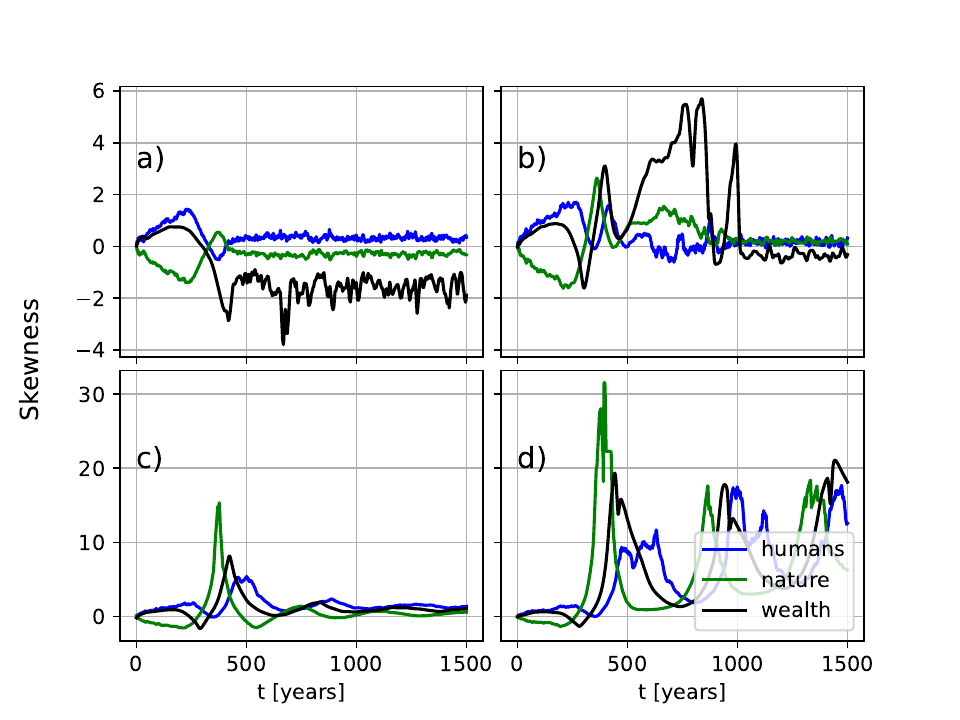} \\
\caption{Skewness associated with state variables Humans, Wealth, and Nature, as a function of time, measured over a set of $1000$ simulations with a random noise of amplitude $\epsilon = 0.03$, for four values of $\delta$, $\delta = \left \{1;  2.5  ; 4 ; 5.5 \right \} \, \delta_{\textnormal{\tiny opt}}$, respectively in the subfigures a-d.}
\label{fig:skew0.03}
\end{figure}

For moderate random noise $\epsilon = 0.03$, the time evolutions of Humans, Nature, and Wealth depart moderately from the standard case, {\it i.e.,} without any random noise.
This is shown in Fig. \ref{fig:skew0.03}, where the skewness is represented as a function of time for the three state variables at play (Humans, Nature, or Wealth), for an ensemble of $1000$ simulations, for $\epsilon = 0.03$, and four values of the depletion factor $\delta = \left \{ 1 ; 2.5 ; 4 ; 5.5 \right \} \delta_{\textnormal{\tiny opt}}$.

At early times of the simulations, typically before $200$ years, for any value of the depletion factor, an anti-correlation is observed between the time dependent skewnesses of Humans and Nature. At such times, the skewnesses associated with Humans as well as Wealth are positive, indicating that these populations tend to exhibit a distribution with a fat tail with respect to the positive values, while the skewness associated with Nature exhibits the opposite behavior, with a fat tail at negative values. This anticorrelation is expected since a positive perturbation of Human Population leads to a faster depletion of Nature, which thus corresponds to Nature’s negative perturbation.
For a depletion factor equal to the optimal value $\delta = \delta_{\textnormal{\tiny opt}}$, Fig. \ref{fig:skew0.03}a, the anticorrelation between Human and Nature skewnesses is preserved throughout the simulated $1500$ model years. The skewness associated with Wealth is positive during the first $200$ model years, while it saturates to a negative value after the transition to quasi-stationary regimes that are observed after $t=500$ years.

For depletion factors larger than the optimal value $\delta = \left \{ 2.5  ; 4 ; 5.5 \right \} \, \delta_{\textnormal{\tiny opt}}$, shown respectively in Figs.  \ref{fig:skew0.03}b-d, the skewness is observed to exhibit high values, mostly positive. In figure \ref{fig:skew0.03}b ($\delta = 2.5 \, \delta_{\textnormal{\tiny opt}}$), the skewness associated with Wealth displays a high and broad peak, between the years $500$ and $1000$, long after the transition to the stationary regime (Fig. \ref{fig:eps0.1}b). For higher depletion factors, oscillations are observed to be rapidly damped for $\delta = 4\, \delta_{\textnormal{\tiny opt}}$, Fig. \ref{fig:skew0.03}c, or to remain important and superimposed to a linear increase for the highest value of depletion considered here, $\delta = 5.5\, \delta_{\textnormal{\tiny opt}}$, Fig. \ref{fig:skew0.03}d.

\begin{figure}[H]
\centering
\includegraphics[scale=1]{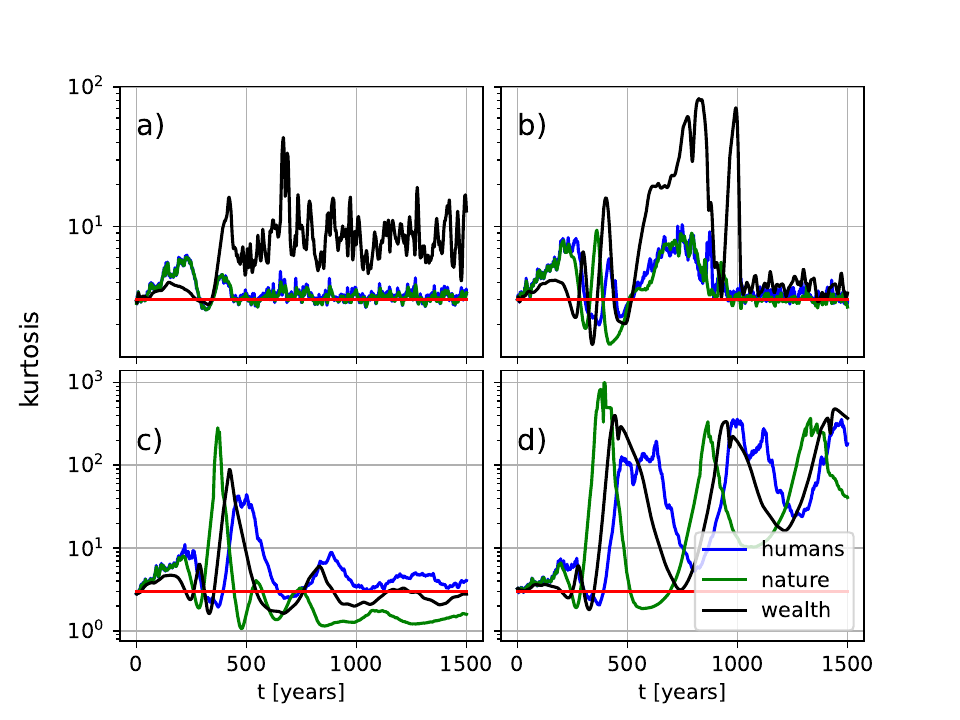} \\
\caption{The kurtosis for the distributions based on 1000 trials is plotted as a function of time for $\epsilon = 0.03$  and for four values of $\delta = \left \{1;  2.5  ; 4 ; 5.5 \right \} \, \delta_{\textnormal{\tiny opt}}$, respectively in the subfigures a-d.}
\label{fig:kurt0.03}
\end{figure}

In Fig. \ref{fig:kurt0.03}, the kurtosises associated with Humans, Nature, and Wealth are represented in logarithmic scale, as functions of time, for the chosen four values of the depletion factor and $\epsilon=0.03$. A red horizontal line is systematically drawn, corresponding to the value associated with a Gaussian distribution of the events.

In most cases, the kurtosis exhibits high values, revealing leptokurtic distributions. The trends observed with respect to the amplitude of the skewnesses are recovered in some sense:
\begin{itemize}
 \item[$\star$] in Fig.  \ref{fig:kurt0.03}a, $\delta =  \delta_{\textnormal{\tiny opt}}$, the evolutions of Humans and Nature tend to exhibit Gaussian distribution, while Wealth displays a higher kurtosis. This reflects the design of the Gaussian perturbation term that is directly applied to Human Population, therefore in its steady state it remains mostly Gaussian. Moreover, Nature, which evolves directly following Population adopts the Gaussian feature of the perturbation. However, Wealth departs from this symmetric Gaussian distribution because it is essentially an accumulation (integral) of production from Nature and becomes more sensitive to a change of those distributions.
 \item[$\star$] in Fig.  \ref{fig:kurt0.03}b, $\delta =  2.5  \delta_{\textnormal{\tiny opt}}$, a transient and broad peak of kurtosis is observed for the Wealth,  related to the same trend for the skewness;
 \item[$\star$] in Fig.  \ref{fig:kurt0.03}c, $\delta = 4  \delta_{\textnormal{\tiny opt}}$, peaks of kurtoses associated with the three state variables are observed during the transient regime, while rapidly damped oscillations lead to long time kurtosis close to Gaussian for Humans and Wealth, and lower than Gaussian for Nature;
 \item[$\star$] in Fig. \ref{fig:kurt0.03}d, $\delta = 5.5 \delta_{\textnormal{\tiny opt}}$, oscillations are observed for the kurtoses of the three state variables, superimposed on a linear growth.
\end{itemize}

The appendix presents a similar study  for $\epsilon=  0.1$. The same observations can be made: for high values of the depletion factor, and with time increasing, the number of collapses becomes so important that the statistical indicators can become irrelevant. It is then appropriate to study the collapse rate as a function of time in order to validate the statistical analysis.

\section{Study of the collapse rate}

Given the high levels of skewness and kurtosis observed, especially for the highest value of the depletion factor $\delta = 5.5\, \delta_{\textnormal{\tiny opt}}$ where a linear growth of the statistical indicators has also been observed, the relevance of these indicators can be questioned. More precisely, taking into account the fact that if at a certain time a quantity is observed to be smaller than a given threshold it is set to zero in order to avoid numerical instability, then it can happen that most of the variables vanish, so that the statistical treatment becomes inappropriate.

\subsection{Time evolution of the collapse rate}

\begin{figure}[H]
\hspace{-3.5cm}
\centering \includegraphics[width=\linewidth]{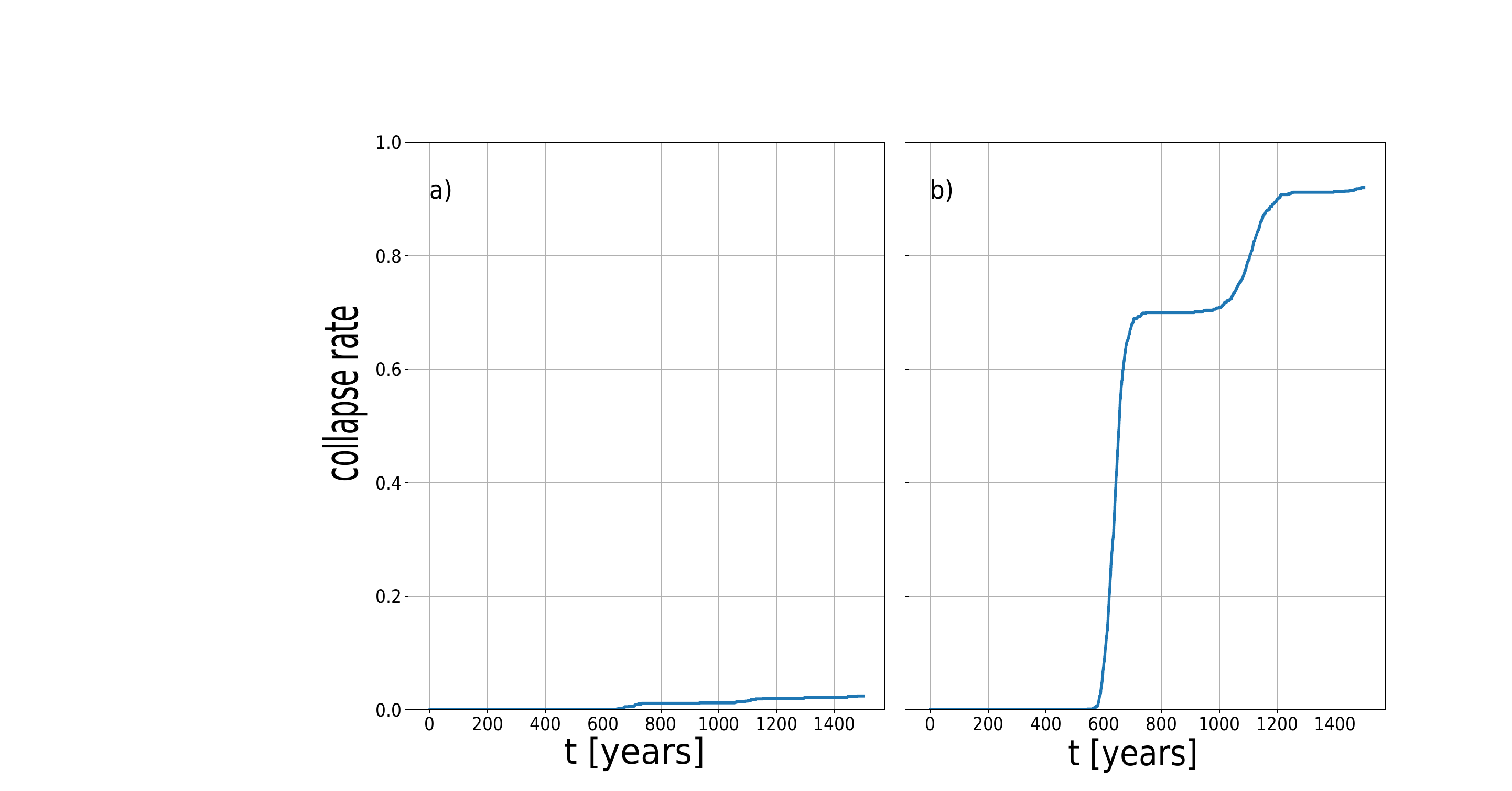} \\
\caption{Fraction of 1000 runs that have collapsed as a function of time for $\epsilon = 0.03$ and for two values of $\delta = \left \{ 4 ; 5.5 \right \} \, \delta_{\textnormal{\tiny opt}}$, respectively in the subfigures a-b. The sharp rise at around $t\simeq 550$ in panel (b) corresponds to the first oscillation of the state variables.}
\label{fig:collapse_time0.03}
\end{figure}

For the calculation of the Collapse Rate, here we define a collapse  as the vanishing of one of the three variables, then forcing the full system to tend to zero. At any time, the collapse rate is defined as the number of collapsed realizations, divided by the total number of realizations ($1000$ in the present case). In Fig.  \ref{fig:collapse_time0.03}, the collapse rate is given as a function of time for the two highest  values of $\delta$. No collapse is observed for the two lowest values of $\delta$, thus they are not presented in the figure. A collapse rate lower than $1\%$ is observed for $\delta = 4 \, \delta_{\textnormal{\tiny opt}}$: the statistical indicators are then relevant for the three sets of simulations associated with the lowest value of $\delta$.

On the other hand, for $t\simeq 550$, the case $\delta = 5.5 \, \delta_{\textnormal{\tiny opt}}$ displays a sudden increase in the collapse rate, reaching $75\%$, and even $90\%$ for  $t\ge 1000$ years. This means that the statistical indicators are not significant here for  $t\ge 500$ years. In particular, the observed linear growth can be related to the huge statistical error introduced by the statistical analysis with a sample that contains more than three fourths of zeros. It should be noted that the increase of the collapse rate a little after  $t=500$ years in the case $\delta = 5.5 \, \delta_{\textnormal{\tiny opt}}$ is located just after  the minimum corresponding to the first oscillation of the variables, as can be seen in subfigure (d) of Fig. \ref{fig:eps0.03}. The action of a random perturbation on the dynamics of variables becomes especially important when the relative size of the perturbation becomes large compared to the value of  the perturbed variable (the Human Population here), which is more likely when the perturbed variable is close to its minimum value.

\begin{figure}[H]
\hspace{-3.5cm}
\includegraphics[scale=0.38]{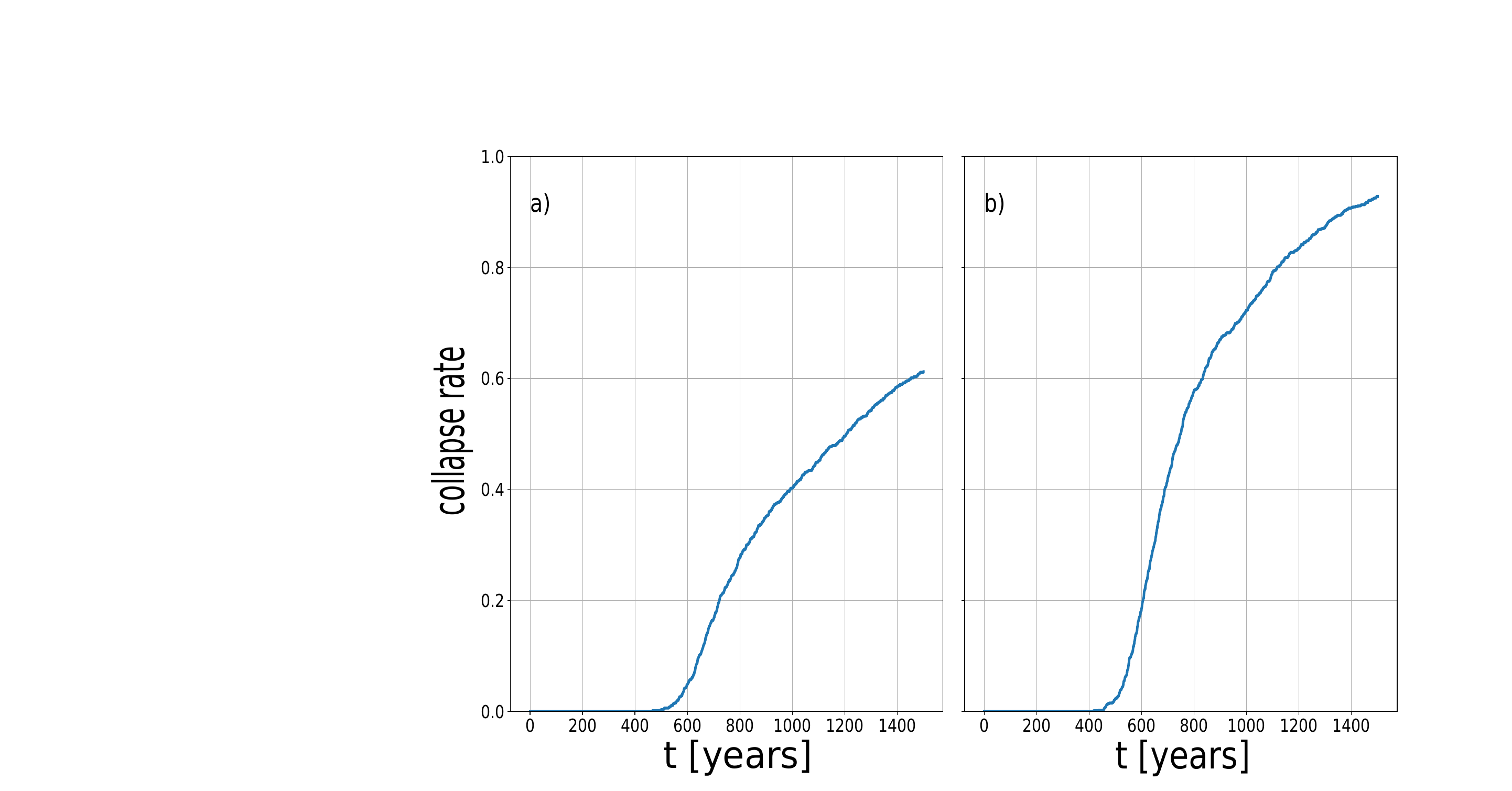} \\
\caption{Fraction of 1000 runs that have collapsed as a function of time for $\epsilon = 0.1$ and for two values of $\delta = \left \{ 4 ; 5.5 \right \} \, \delta_{\textnormal{\tiny opt}}$, respectively in subfigures a-b}
\label{fig:collapse_time0.1}
\end{figure}

In Fig. \ref{fig:collapse_time0.1}, the time variation of the collapse rate is given for the two
values of the depletion factor $\delta = \left \{ 4 ; 5.5 \right \} \, \delta_{\textnormal{\tiny opt}}$ and $\epsilon = 0.1$. For high depletion factors $\delta \geq 4 \delta_{\textnormal{\tiny opt}}$, high levels of collapse rates are observed: over 40\% or 60\% of the realizations collapse after 1000 years. As discussed in the case $\epsilon = 0.03$, such an observation makes the analysis of statistical indicators not adequate. More importantly, the increase of the collapse rate starts before  $t < 500$ years. This $t=500$ years still corresponds to the first minimum values of the state variables (Fig.  \ref{fig:eps0.03}). The year at which it occurs can also be linked to the amplitude of the random noise and to the strength of the depletion rate. The higher these parameters, i.e., $\delta$ and $\epsilon$, the sooner the collapse occurs.

\subsection{Final state as a function of the amplitude of perturbation, $\epsilon$}

In the previous sections, the collapse rate has been analyzed as a function of time for fourty values of the amplitude of the statistical noise $\epsilon$. This indicator and its dependence with respect to the amplitude of the random noise is examined in detail in the present section.

\begin{figure}[H]
\includegraphics[scale=1]{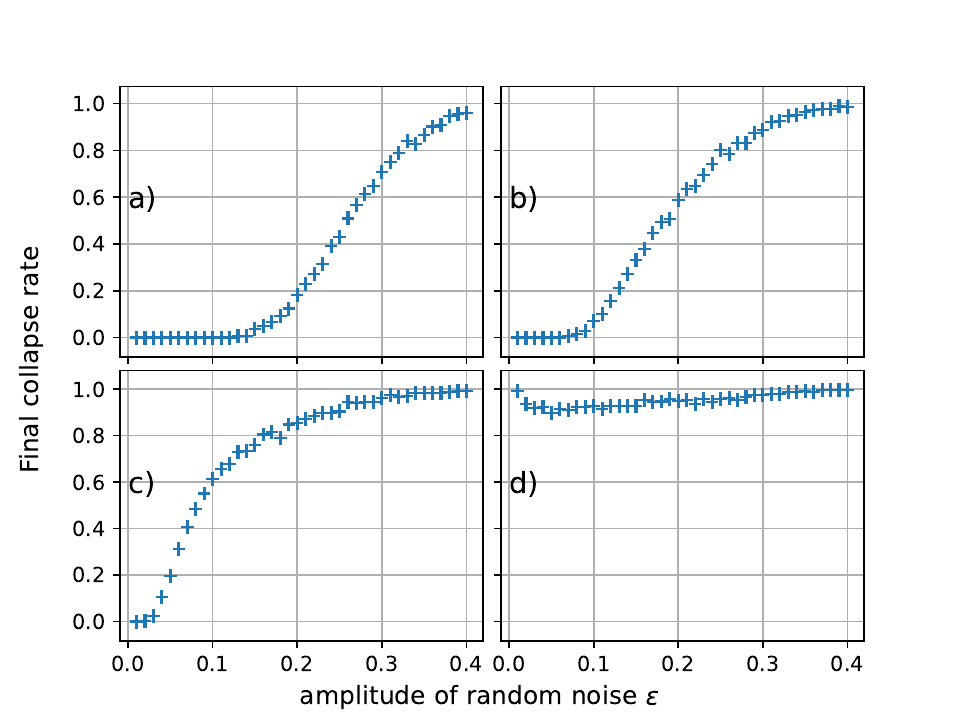} \\
\caption{Final collapse rate as a function of the amplitude of the random noise, for four values of the departure from optimum $\delta = \left \{ 1 ; 2.5 ; 4 ; 5.5 \right \} \delta_{opt}$. }
\label{fig:collapse_rate}
\end{figure}

In Fig. \ref{fig:collapse_rate}, the collapse rate measured at the end of the simulations ($t = 1 500$ years), is shown as a function of the amplitude of the random noise for the four values of the depletion factor $\delta = \left \{ 1 ; 2.5 ; 4 ; 5.5 \right \} \delta_{\textnormal{\tiny opt}}$.

For an optimal value of the depletion factor $\delta = \delta_{\textnormal{\tiny opt}}$, Fig \ref{fig:collapse_rate}a, the collapse rate increases  with respect to the amplitude of the random noise $\epsilon$. Going further into details, the variation is found to follow an hyperbolic tangent shape, with a collapse rate equal to one half for $\epsilon \approx 0.25$. The increase is sharp: below $\epsilon \lesssim 0.18$, the collapse rate is lower than $10$\%, while it becomes larger than $75 $\% for $\epsilon \gtrsim 0.3$.

Increasing the depletion factor to $\delta = 2.5 \delta_{\textnormal{\tiny opt}}$, Fig. \ref{fig:collapse_rate}b, the hyperbolic tangent shape is recovered, but with a broader range for the transition between low collapse rate and high collapse rate regions. In particular, it is associated with a narrower range of the region with low collapse rate: the threshold for increasing the collapse rate over  $10$ \%,  extends from $\epsilon \approx 0.18$ (Fig. \ref{fig:collapse_rate}a, optimal $\delta$), to $\epsilon \approx 0.1$ in the case of $\delta = 2.5 \delta_{\textnormal{\tiny opt}}$ (Fig. \ref{fig:collapse_rate}b).

With a higher depletion factor $\delta = 4 \delta_{\textnormal{\tiny opt}}$, Fig. \ref{fig:collapse_rate}c, the hyperbolic tangent shape appears to shift to a natural logarithm at $\epsilon = 0.1$. The region with almost no collapse is notably limited to values $\epsilon \lesssim 0.03$.

For the highest depletion factor considered, $\delta = 5.5 \delta_{\textnormal{\tiny opt}}$, Fig. \ref{fig:collapse_rate}d, the collapse rate is higher than $80$\% regardless of the amplitude of the random noise. One can note the existence of a region with a lower collapse rate for $\epsilon \in \left [ 0.01 ; 0.3 \right ]$, with minimal values around $\epsilon \approx 0.95$. However, such values remain high and most of the realizations have collapsed. Qualitatively speaking, the fact that some of the simulations have not yet collapsed after $1500$ years is due to the random nature of the noise, together with the low values of the human population compared to the high amplitude of the random perturbation.

\begin{figure}[ht]
    \centering 
    
    \includegraphics[width=\linewidth,clip=true,trim=14cm 2cm 6cm 5cm]{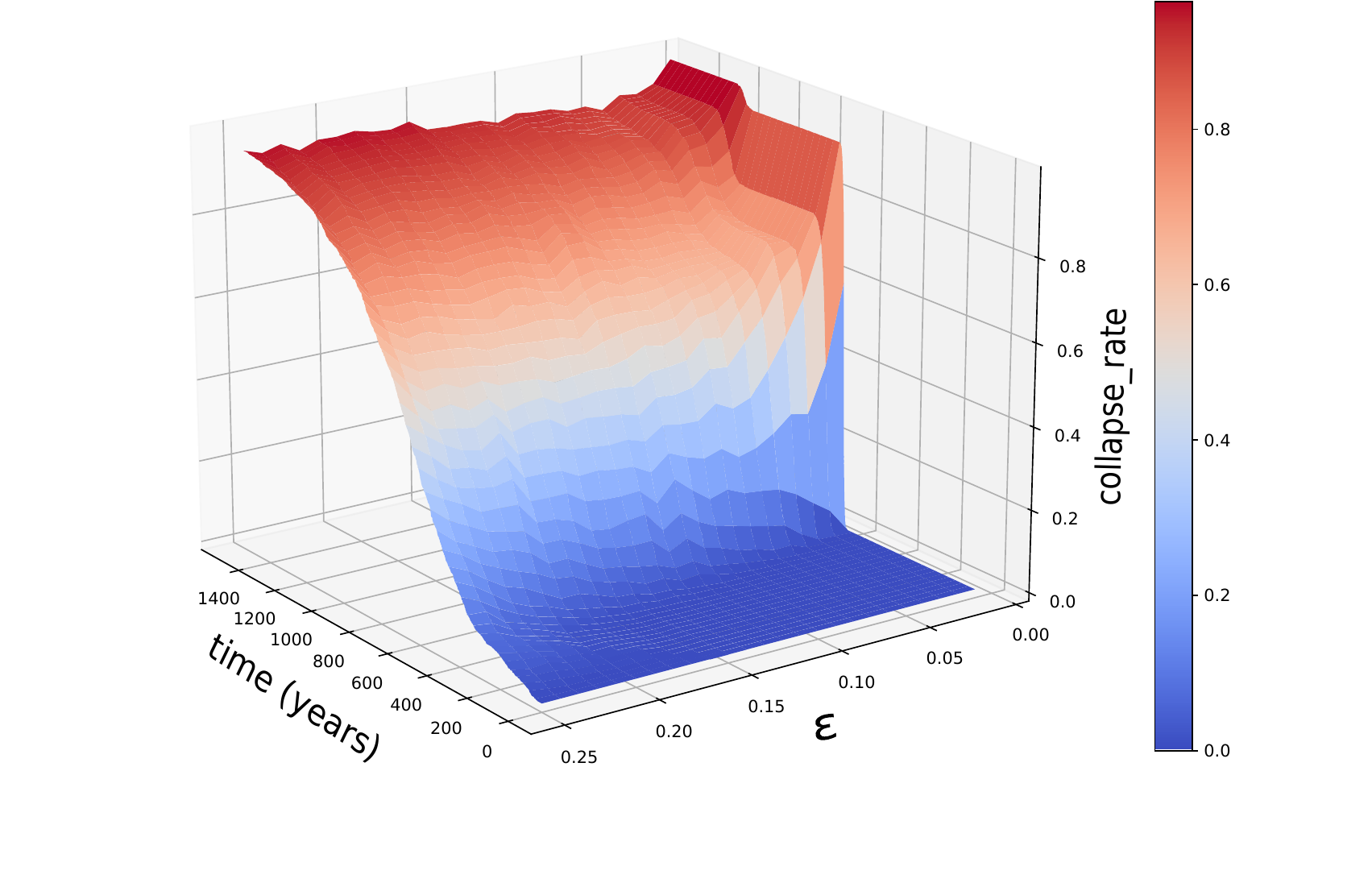}
    \caption{Fraction of 1000 runs that have collapsed as a function of time ($t$ in years), and of the random noise amplitude $\epsilon$, for a maximal depletion factor $\delta = 5.5 \, \delta_{\textnormal{\tiny opt}}$, showing a sudden (smooth)  increase at small (large) $\epsilon$.}
    \label{fig:collapse_rate_2D}
\end{figure}

The comparison of   figures \ref{fig:collapse_time0.03}-d and \ref{fig:collapse_time0.1}-d corresponding to the highest depletion factor $\delta = 5.5 \delta_{\textnormal{\tiny opt}}$ reveals a transition between a multiple steps with respect to time behavior, to a smooth behavior, while increasing the amplitude of the random noise. This phase transition is shown  in Fig. \ref{fig:collapse_rate_2D}, where the collapse rate is shown as a function of both $\epsilon$ and time.

While varying the amplitude of the random noise $\epsilon$, the step-like behavior of the collapse rate is progressively lost and replaced by a monotonic increase with respect to time. The transition between these two regimes occurs around random noise amplitudes of $\epsilon_c \approx 0.10$.

To summarize this section, the collapse rate is found to increase with the amplitude of the random noise as well as with the depletion factor. For the highest value of the depletion factor $\delta = 5.5 \delta_{\textnormal{\tiny opt}}$, a transition is found between sudden increases at low random noise amplitudes, and smooth increases at higher amplitudes of the random noise.

\section{Conclusion}

The HANDY model is based on the predator--prey (Lotka-Volterra) and logistic systems. It describes the evolution of the human population, separated into Commoners and Elites, Nature, and accumulated Wealth. For simplicity, we restricted our study to egalitarian society with only one type of population.
We added random noise to HANDY to investigate its robustness under random perturbations. We carried out statistical analysis based on probability distributions. Random perturbations are introduced in the form of events of different amplitudes drawn from Gaussian distributions, and their effects on the final state of the system are analyzed. When random perturbations are introduced in the dynamics of Human Population, they model what may happen during pandemics, wars, earthquakes, floods, droughts, or in general by any sudden change of the human population.

A large number of trials were carried out in order to analyze the second (standard deviation), third (skewness), and fourth (Kurtosis) moments of the obtained distributions.

Our study shows that the results of the unperturbed HANDY model are robust under small  perturbations of $\lesssim$ 3\% of the Human population, i.e., the endogenous interactions of the Human--Nature system determines its sustainability or collapse. Nevertheless, perturbations can hasten or delay a collapse cycle.

The non-linear structure of the dynamical system explains the departure from gaussianity of the response of the HANDY system to Gaussian random perturbations. If the amplitude of the Gaussian random perturbations becomes large, then even a scenario with a stable equilibrium can be pushed to collapse. This is expected because in this case the perturbation dominates the dynamics of the system. We note that such large perturbations are extremely rare in the real world. We investigated them from a mathematical perspective and found results consistent with our intuition about this dynamical system.  Our conclusions confirm the  qualitative dynamics of HANDY, in line with the original paper \cite{MOTESHARREI201490}.

\section{Acknowledgments}

The interest of two of the authors (P.M. and E.T-G.) to the subject was driven by the work of Pablo Servigne and Rapha\"el Stevens "Comment tout peut s'effondrer". We are grateful to Professor James Yorke for interesting discussions and valuable suggestions. E.K. and S.M. acknowledge the generous support from Drs. Eugenia and Michael Brin. SM also acknowledges the National Science Foundation RTG grant DMS-2136228.

\section{Appendix : Departure from gaussianity for $\epsilon = 0.1$ \label{sec:appendix}}

Additional results of skewness and kurtosis are reported here for the case of $\epsilon=0.3$.

\begin{figure}[H]
\centering
\includegraphics[scale=1]{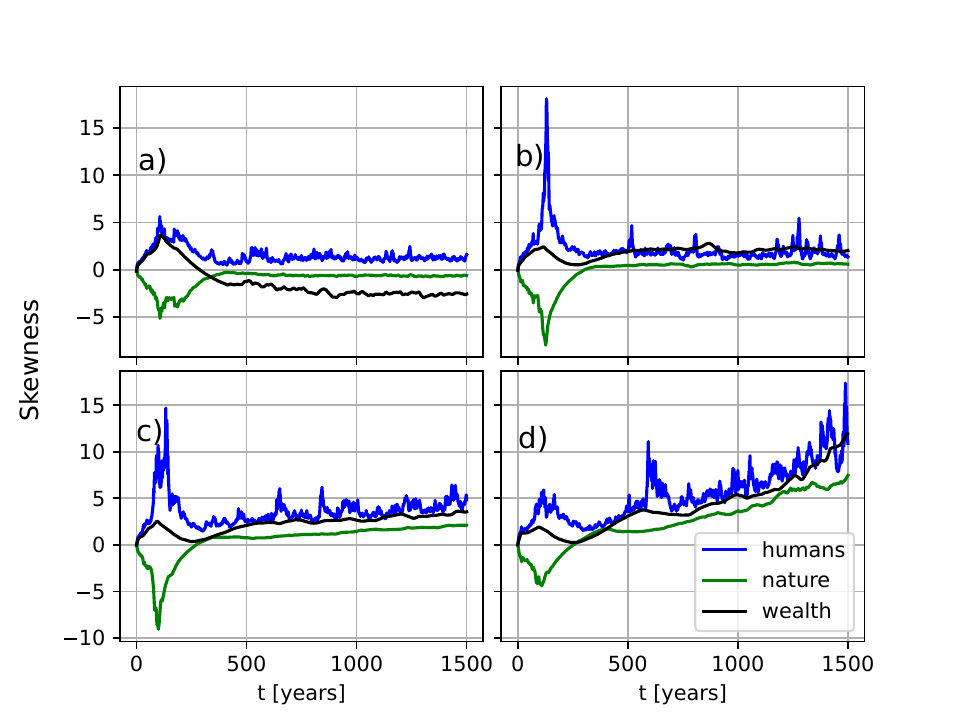} \\
\caption{Skewness associated with Humans, Nature, and Wealth  as a function of time, measured over a set of $1000$ simulations with a random noise of amplitude $\epsilon = 0.1$, for four values of $\delta = \left \{1;  2.5  ; 4 ; 5.5 \right \} \, \delta_{\textnormal{\tiny opt}}$, respectively, in subfigures a--d.}
\label{fig:skew0.1}
\end{figure}

When the amplitude of the noise is increased to 10\% of the main signal, $\epsilon = 0.1$, the changes in the observed behaviors with respect to the case without noise become significant: the oscillatory phases are damped and the two set of simulations with the highest depletion factors collapse systematically (Fig. \ref{fig:eps0.1}). The standard deviations of most distributions become very large, losing the Gaussian feature.

\begin{figure}[H]
\centering
\includegraphics[scale=1]{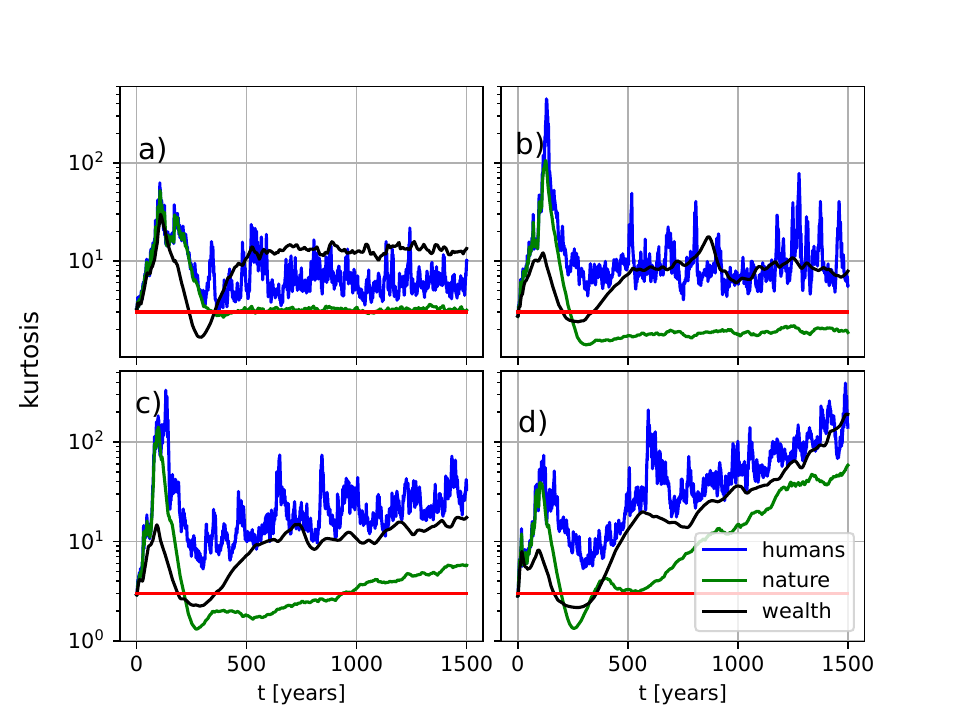} \\
\caption{The kurtosis for the distributions based on 1000 trials is plotted as a function of time for $\epsilon = 0.1$  and for four values of $\delta = \left \{1;  2.5  ; 4 ; 5.5 \right \} \, \delta_{\textnormal{\tiny opt}}$, in the subfigures a--d, respectively.}
\label{fig:kurt0.1}
\end{figure}

One observes a correlated behavior for skewness and kurtosis. For optimum $\delta$, the skewness is about Gaussian for Humans and Nature, Fig. \ref{fig:skew0.1}a. For $t \simeq 1000$ years, a bump in the skewness of Wealth is observed, which correlates with a large bump in the associated kurtosis Fig. \ref{fig:kurt0.1}a. Meanwhile, the indicators associated with Nature and Humans are not affected and seem to reach a steady-state.

Increasing $\delta$ to $\delta = 2.5 \delta_{\textnormal{\tiny opt}}$, steady-state may still be reached, Figs. \ref{fig:skew0.1}b and \ref{fig:kurt0.1}b, with the probability of extreme and positive events of Nature and Wealth distributions being larger than a Gaussian, while the distribution for Nature is observed to be symmetric and platykurtic, {\it i.e.,} with less extreme events than in the case of the Gaussian distribution.

Further increase of $\delta$ to $\delta = 4 \delta_{\textnormal{\tiny opt}}$ or $\delta = 5.5 \delta_{\textnormal{\tiny opt}}$ prevents the system from reaching a steady-state (Fig. \ref{fig:eps0.1}). All distributions depart significantly from a Gaussian distribution, making statistical analysis unreliable, as can be observed from Figs. \ref{fig:skew0.1}c-d, \ref{fig:kurt0.1}c-d.

For the weight of the statistical noise considered here $\epsilon = 0.1$, the kurtosis and skewness are observed to present a secular linear growth associated with values of the depletion factor $\delta > 4 \delta_{\textnormal{\tiny opt}}$.


\end{document}